\documentclass[epj]{webofc}
\usepackage[utf8]{inputenc}
\usepackage[varg]{txfonts}   
\usepackage{booktabs,ulem}
\usepackage{xcolor}
\definecolor{darkred}{rgb}{0.4,0.0,0.0}
\definecolor{darkgreen}{rgb}{0.0,0.4,0.0}
\definecolor{darkblue}{rgb}{0.0,0.0,0.4}
\usepackage[bookmarks,linktocpage,colorlinks,
    linkcolor = darkred,
    urlcolor  = darkblue,
    citecolor = darkgreen]{hyperref}
\wocname{EPJ Web of Conferences}
\woctitle{Lattice2017}
\begin{document}
%
\selectlanguage{english}
\title{%
Testing Fermion Universality at a Conformal Fixed Point
}
\author{%
\firstname{Anna} \lastname{Hasenfratz}\inst{1}\fnsep\thanks{Speaker, \email{Anna.Hasenfratz@colorado.edu }} \and
\firstname{Claudio} \lastname{Rebbi}\inst{2} \and
\firstname{Oliver}  \lastname{Witzel}\inst{3}
}
\institute{%
Department of Physics, University of Colorado, Boulder, CO 80309, USA
\and
Department of Physics and Center for Computational Science, Boston University, Boston, MA 02215, USA
\and
Higgs Centre for Theoretical Physics, University of Edinburgh, Edinburgh, EH9 3FD, UK
}
\abstract{%
 Universality of various fermion formulations is well established in QCD-like theories defined around  the perturbative $g^2=0$ fixed point. These arguments do not 
apply  for conformal systems that exhibit an infrared fixed point at non-vanishing $g^2$ coupling. 
We investigate the step scaling function for  systems with 10 or 12 fundamental flavors using domain wall fermions and compare it to perturbative predictions. We test universality by contrasting our findings to results published in the literature based on staggered fermions.
}
\maketitle
\section{Introduction}\label{intro}

The concept of universality is a cornerstone of quantum field theory investigations on the lattice. 
Universality means that systems with identical field content and identical symmetries in the same space-time dimension have the same universal critical properties. Since the continuum limit is reached when the system is tuned to criticality, universality implies that the continuum limit depends only on  the relevant operators of the system.   We point out  that a unique continuum limit can be defined in the basin of attraction of every ultraviolet fixed point (UVFP) of a system. 

The symmetries relevant for universality include local gauge invariance, and global ones like  flavor symmetry of fermions.  While local gauge symmetries are usually preserved by the lattice action, the flavor symmetries are frequently not. Staggered fermions break the $\text{SU}(N_f)\times \text{SU}(N_f)$ symmetry to $\text{SU}(N_f/4)\times \text{SU}(N_f/4)$, while Wilson fermions break flavor symmetry completely. Thus  universality arguments do not apply to fermions unless the flavor symmetry gets restored at criticality. 

QCD-like gauge-fermion systems  are asymptotically free and the continuum limit is defined around the perturbative $g^2=0$ Gaussian fixed point (GFP) where the restoration of continuum flavor symmetry can be proven perturbatively.  In case of staggered fermions the taste breaking terms enter as $\mathcal{O}(g^2)$.  Proving  the continuum flavor symmetries are recovered as the bare gauge coupling $g^2$ is tuned to zero is  non-trivial,  because one has to show that all taste-breaking operators are irrelevant at the GFP~\cite{Golterman:1984cy}. (For a concise summary see  e.g.~Ref.~\cite{Sharpe:2006re}.) 
 If one increases the number of flavors,  a new fixed point, in addition to the GFP, emerges where the gauge coupling becomes  irrelevant.  Even though this fixed point is commonly referred to as  conformal infrared fixed point (IRFP), it is actually an UVFP in the mass, which is the only relevant operator.  The continuum limit in the basin of attraction of  this conformal FP is reached  by tuning the fermion mass  to zero. The gauge coupling does not require tuning; in the infrared it takes the value of the IRFP, independent of its bare value at the cut-off.  The proof showing staggered fermions are in the same universality class as continuum fermions \cite{Golterman:1984cy} relies however on $g^2 \to 0$. Thus if the taste breaking terms of staggered fermions do not vanish at the conformal IRFP, staggered formulations of a conformal system might not be in the same universality class as continuum fermions.\footnote{This issue is not related to rooting. Whether a rooted staggered action is equivalent to a local action at an IRFP  with $g^2>0$ is  another interesting question, but  not considered here.}

A situation similar to the taste breaking of staggered fermions can be modeled in the 3-dimensional $O(n)$ scalar model. For illustration  we summarize the results of Ref.~\cite{Kleinert:1994td}  in the next section before highlighting recent developments for SU(3) gauge theories in  four dimensions. 

In Sect.~\ref{sec:4D} we  present our new results on the  step scaling functions for SU(3) gauge theories with  10 or 12 fundamental flavors  obtained with domain wall fermions (DWF).   Our findings  suggest that at a conformal IRFP  staggered fermions are not in the same universality class as  domain wall fermions. Finally, we conclude in Sec.~\ref{sec:conclusion}.

\section{Indications of universality violation }
\subsection{\texorpdfstring{$\phi^4$  scalar models in three dimensions}{phi4  scalar models in three dimensions}}\label{sec:3D}
%
\begin{figure}[thb] 
  \centering
  \includegraphics[width=0.40\columnwidth,clip]{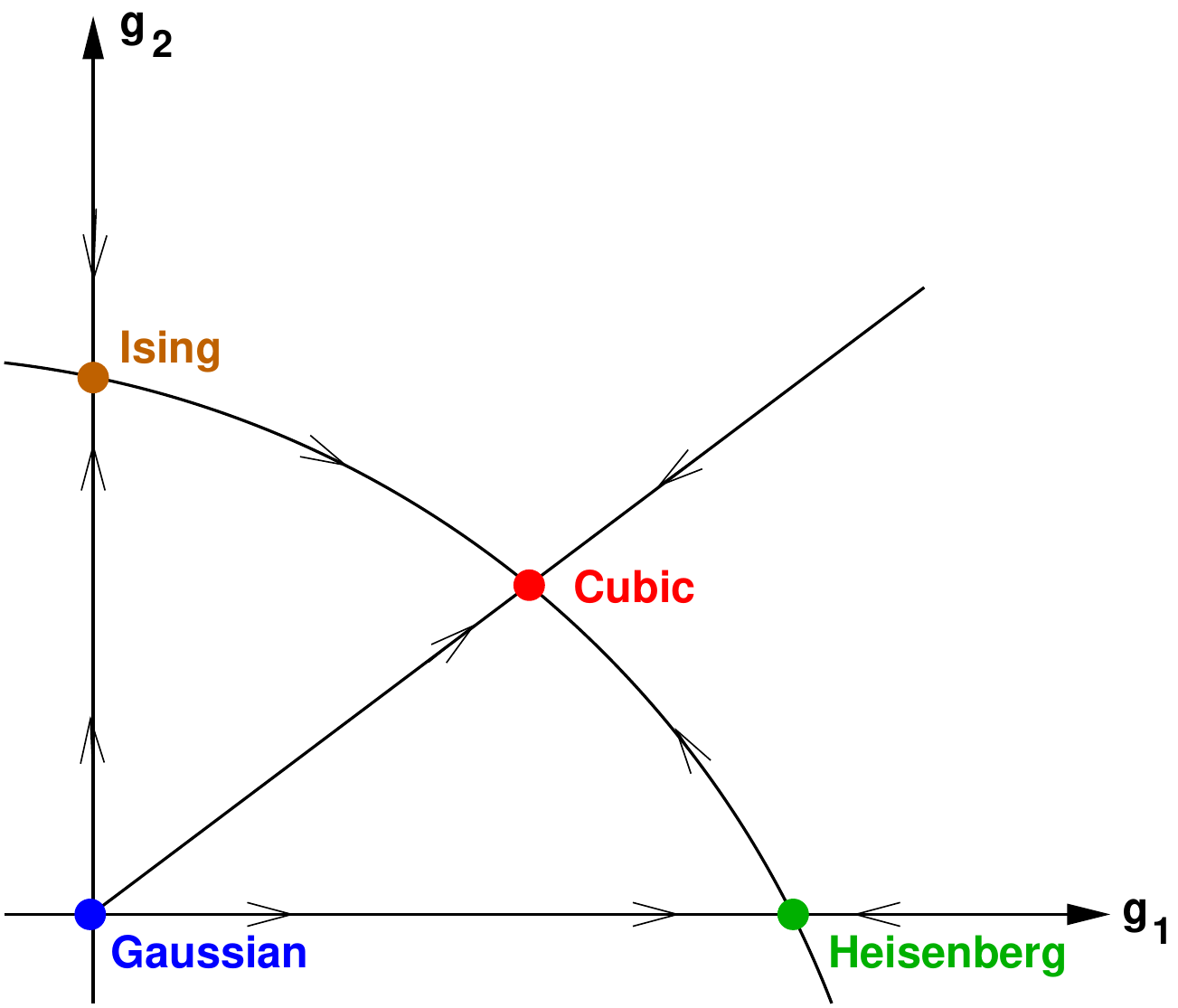}
  \caption{Sketch of the RG flows on the critical surface of the scalar model with symmetry breaking term as  in Eq.~(\ref{eq:Kleinert}). The infrared limit is characterized by a new "cubic" fixed point~\cite{Kleinert:1994td}.  }
  \label{fig:scalar}
\end{figure}

The $O(n)$ symmetric  scalar models in  three dimensions have  very similar structure to the 4-dimensional conformal gauge-fermion  systems.  The $\phi^4$ coupling is relevant at the perturbative GFP but irrelevant at the non-perturbative conformal Wilson-Fischer fixed point (WFFP), whereas the mass is a relevant operator at both fixed points. The GFP has mean-field exponents for all $n$, but the WFFP  is different  for different $n$. Adding to the action interaction terms breaking the symmetry  $O(n) \to O(n^\prime)$,  multicritical phenomena can emerge (see e.g. Refs.~\cite{Kleinert:1994td,Calabrese:2002bm}).   Using the $\epsilon$-expansion  up to fifth order  a  model with potential
\begin{equation}
  \label{eq:Kleinert}
 V = \frac{1}{2} m^2 \left(\sum_{\alpha=1}^n \phi_n^2\right)  + g_1 \left(\sum_{\alpha=1}^n \phi_n^2\right)^2 + g_2 \left(\sum_{\alpha=1}^n \phi_a^4\right),
\end{equation}
 is investigated in   Ref.~\cite{Kleinert:1994td}. In addition to the perturbative GFP, the system has  an $O(n)$ Heisenberg WFFP for $g_2=0$ and a $Z_2$ symmetric Ising FP  for $g_1=0$.
 The renormalization group structure of this system when both couplings are non-zero is similar to the sketch in Fig.~\ref{fig:scalar}. The infrared limit is governed by neither by the $O(n)$ nor the $Z_2$ symmetric fixed point but by a new, ``cubic'' fixed point.  Perhaps the situation is similar for staggered fermions in 4-dimensional conformal models.

\subsection{SU(3) gauge theory in four dimensions}\label{sec:motivation} 
Recently the gradient flow step scaling function has been used to investigate the $\beta$ function in 4-dimensional SU(3) gauge theories~\cite{Luscher:2010iy,Fodor:2012td}. The gradient flow step scaling function for a scale change $s$ is related to the discrete $\beta$-function 
\begin{equation}
  \label{eq:beta_L}
  \beta_s(g^2_c; L) = \frac{g^2_c(sL; a) - g^2_c(L; a)}{\log(s^2)}.
\end{equation}
The ratio $c=\sqrt{8t}/L$ ties the lattice volume to the energy scale where the parameter $t$ is the flow time,  $L$  the extent of the lattice with volume $L^4$, and $g_c^2(L ; a)$ the gradient flow coupling  at lattice spacing $a$.~\footnote{Following the usual conventions,  $t$, $L$, and $a$ are dimensionful quantities.}  The continuum extrapolated discrete $\beta$-function $\beta_s(g^2_c) = \lim_{(a / L) \to 0} \beta_s(g^2_c, L)$ depends only on the renormalized coupling $g^2_c$, therefore it is expected to be independent of  irrelevant operators introduced by the lattice regularization. The gradient flow step scaling function is only 1-loop universal when the simulations are done with periodic or antiperiodic boundary conditions. We do not expect the non-perturbative lattice results to follow  perturbative curves outside the small-$g^2_c$ range.  In the plots we nevertheless show  the perturbative 2-loop and 4-loop $\overline{MS}$ predictions. These solely serve  to guide the eye and help to compare results obtained with different lattice actions.
 In the remainder of this section we briefly point out two cases possibly indicating violations of universality which motivated our investigations:
\begin{enumerate}
\item \textbf{SU(3) gauge model with  2 flavors in the sextet representation:} Using Wilson fermions and the Schr{\"o}dinger functional scheme Refs.~\cite{Shamir:2008pb,DeGrand:2013uha} found  the step scaling function to be consistently smaller than the 2-loop value, possibly developing an IRFP.  More recently Ref.~\cite{Hasenfratz:2015ssa} investigated the $c=0.35$ gradient flow step scaling function with Wilson fermions and  found that it approximately follows the 4-loop $\overline{MS}$ prediction in the $0<g^2_c <5.5$ range, increasing for $g^2_c\lesssim 3.5$, decreasing thereafter and  approaching zero around $g^2_c \approx 6.0$.
The Lattice Higgs Collaboration studied the gradient flow step scaling function in the same scheme using (rooted) staggered fermions.  Their results predict that $\beta_s(g^2_c)$ increases monotonically in the range $0 <  g^2  <  6.5$, staying within $\approx 20\%$ of the 2-loop value \cite{Fodor:2015zna}.
 Establishing a zero in the step scaling function is difficult, however the qualitative differences between staggered and Wilson results are striking and are too large in the $g^2_c \ge 4.0$ range   to be attributed to under-estimated statistical errors.

\item \textbf{SU(3) gauge theory with $N_f\gtrsim 8$ fundamental flavors:} 2-loop perturbation theory predicts  the boundary of the conformal window to be at $N_f \gtrsim 8$. The step scaling function of the $N_f=12$ flavor system has been investigated extensively with staggered fermions. Figure 5 of  Ref.~\cite{Hasenfratz:2016dou}, reproduced on the right panel of Fig.~\ref{fig:Nf12-DW}, shows results from three different calculations that agree within errors and predict the step scaling function to lie between the 2- and 4-loop perturbative values with a suggested IRFP around $g^2_c\approx 7.4(3)$.\footnote{ At Lattice 2017 the LatHC collaboration presented a poster with updated results indicating that the step scaling function in the $g^2_c \approx 7.0-7.6$  range is close to 0.1, i.e. it has minimal dependence on the gauge coupling in a wide $g^2_c$ range. In this paper we concentrate on the overall  shape  of $\beta_s(g^2_c)$ at smaller couplings.}  This result is in some tension  with a recent domain wall fermion calculation  of a 10-flavor system that predicts the gradient flow step scaling function well below the 4-loop $\overline{MS}$ curve and suggests an IRFP around $g^2_c=7.0$~\cite{Chiu:2016uui,Chiu:2017kza}.  In general $N_f=10$ is expected to show features  that are related to the conformal window, like a real or approximate IRFP,  at stronger coupling than $N_f=12$. The qualitative features of the step scaling functions with staggered and DW fermions appear to be in conflict with each other.
\end{enumerate}

 \section{Step scaling function from domain wall simulations}\label{sec:4D}
\begin{figure}[thb] 
  \centering
  \includegraphics[width=0.49\columnwidth,clip]{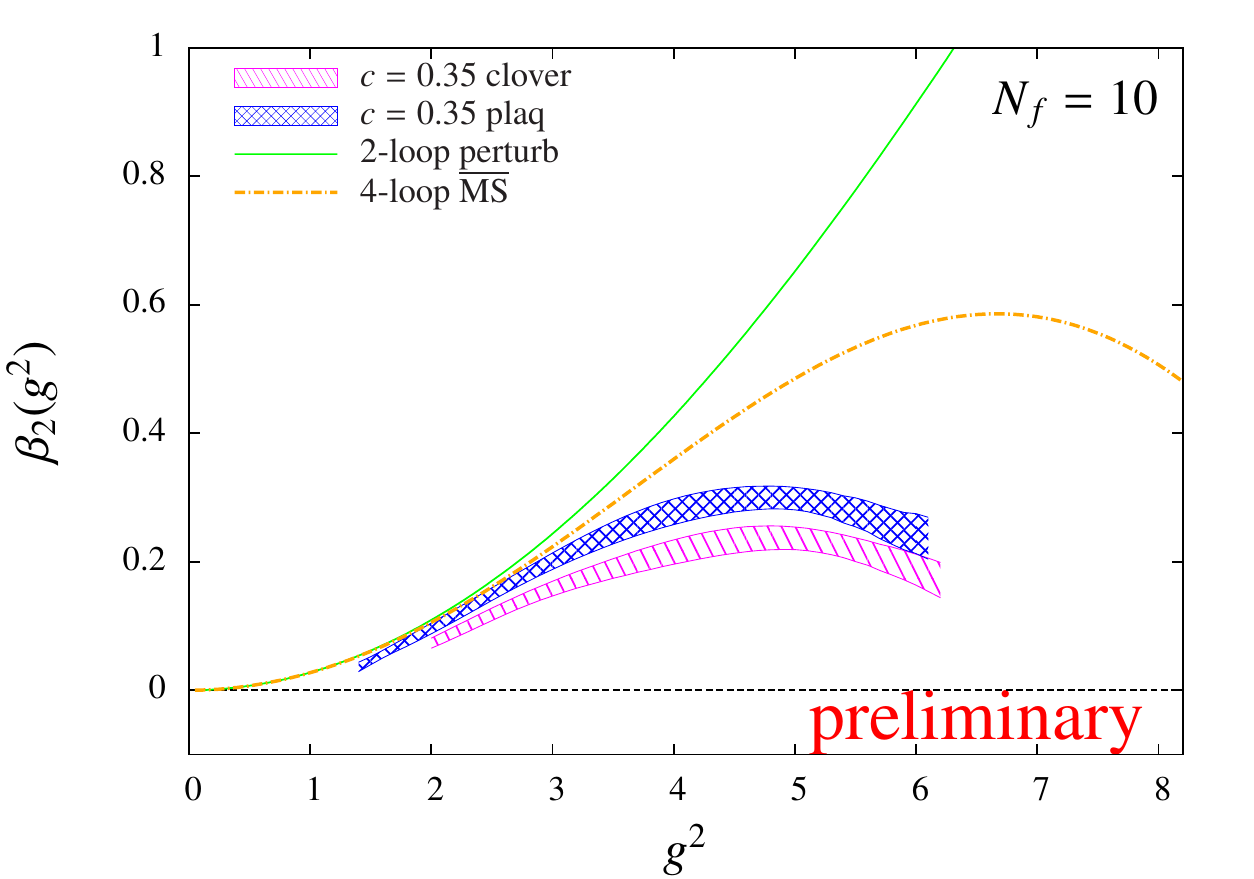}
  \includegraphics[width=0.49\columnwidth,clip]{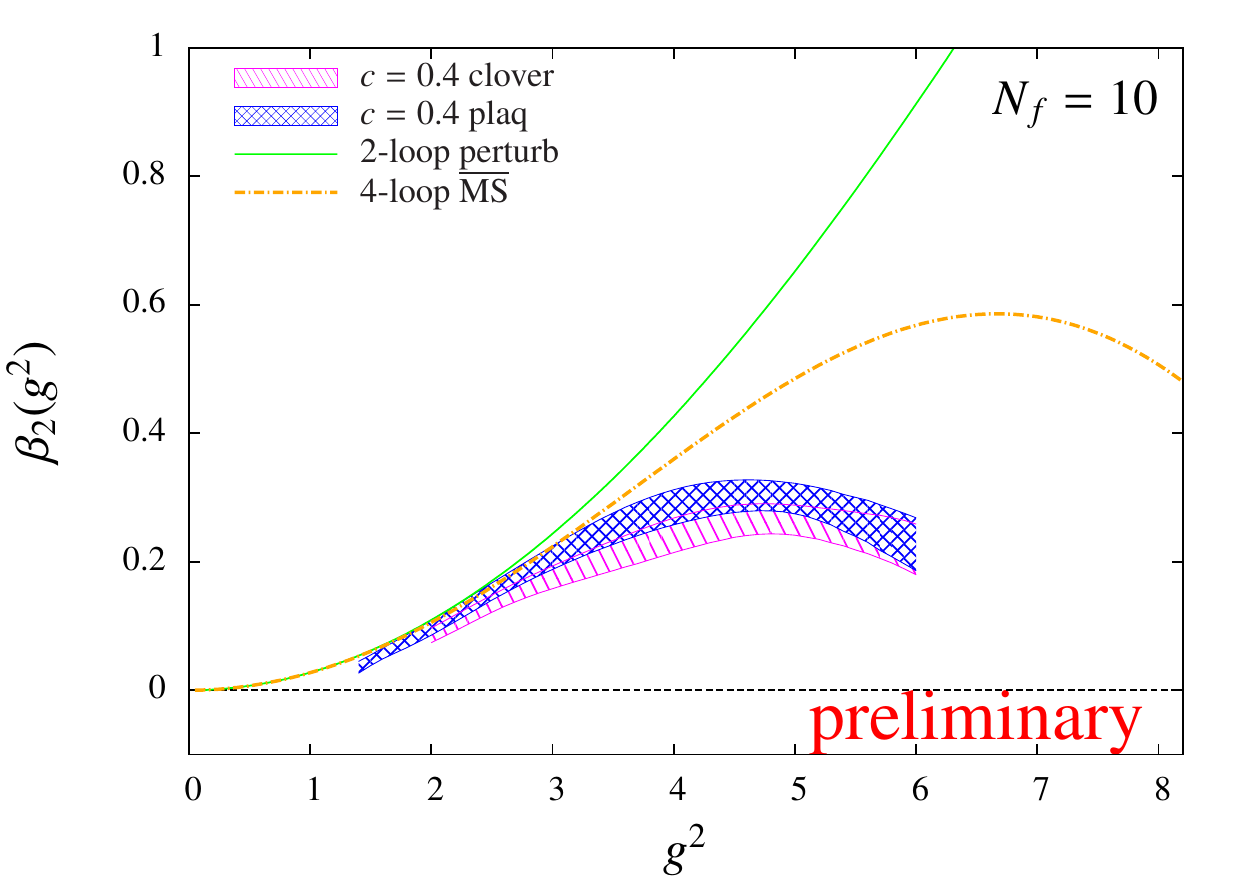}
  \caption{Comparison of the  10-flavor $(a/L)^2 \to 0$ extrapolated step scaling functions using the plaquette and clover discretizations  in our MDW fermion simulations. Left panel corresponds to $c=0.35$, right to $c=0.4$.  }
  \label{fig:Nf10-DW-compare}
\end{figure}
 
Motivated by the possible lack of universality of staggered fermions at a conformal fixed point we have started a program to investigate the step scaling function of SU(3) gauge theory with $N_f=10$ and 12 fundamental flavors. We use M{\"o}bius domain wall fermions  with 3-levels of stout smearing and Symanzik gauge action~\cite{Brower:2012vk,Morningstar:2003gk,Kaneko:2013jla,Noaki:2015xpx}.  The boundary conditions are periodic  for the gauge fields and antiperiodic for the fermions in all  four directions and the bare fermion mass is set to zero. Our simulations are carried out using the \texttt{Grid} code \cite{Boyle:2015tjk,GRID}.

 We simulate symmetric $L^4$ volumes with $L/a=6$, 8, 10, 12, 16, 20 and 24, although in our analysis we use only $L/a \ge 8$. On all seven volumes we have generated  ensembles at 12 -- 14 values of the gauge coupling in the range $4.15 \le \beta \le 7.0$ for $N_f=10$ and $4.20\le \beta\le 7.0$ for $N_f=12$.  For each ensemble  we have typically collected 5000  Molecular Dynamics  Time Units (MDTU). We identified a first order bulk transition for both  $N_f=10$ and 12 around $\beta=4.0 - 4.05$. The DW residual mass increases exponentially as this transition is approached, thus limiting the range of couplings we can simulate.   

For most ensembles we chose the 5th dimension of domain wall  fermions to be $L_s=12$,  but increase it up to $L_s=24$ on the larger ($L/a=16,\,20,\,24$) volumes at  stronger gauge coupling in order to keep the residual mass sufficiently small.  If the residual mass is too large compared to the energy scale given by the inverse lattice size $L^{-1}$, the configurations should be considered finite-mass deformed and not volume squeezed. At finite fermion mass the gauge coupling runs faster and the step scaling function predicted by the finite-volume gradient flow scheme would be larger than in the chiral limit. In this sense our results here can be considered as upper bounds  on $\beta_s(g^2_c)$.

The volumes in our simulations are small compared to staggered fermion calculations. We expect  that lattice artifacts with DW fermions are smaller than with staggered, as many QCD simulations indicate.  The DW calculation of the step scaling function of  Refs.~\cite{Chiu:2016uui,Chiu:2017kza} also indicates that $\beta_s(L;g^2_c)$ shows linear   $(a/L)^2$ dependence already at $L/a=8$. We have started simulations with $L/a=32$  to strengthen the $(a/L)^2 \to 0$ extrapolation, but they are not included here. We measure the gradient flow renormalized coupling using Wilson flow and consider both the clover and plaquette discretization of the energy density. We choose $c=\sqrt{8t}/L$ large enough so the two are consistent.   We also apply $t$-shift optimization to remove $\mathcal{O}(a^2)$ cut-off effects~\cite{Hasenfratz:2015xpa} and plan to repeat the gradient flow measurements with Symanzik flow as a further consistency check. 

\subsection{\texorpdfstring{Step scaling function with $N_f=10$ fundamental flavors}{Step scaling function with Nf=10 fundamental flavors}}\label{sec-Nf10}
%

\begin{figure}[thb] 
  \centering
  \includegraphics[height=0.2\textheight]{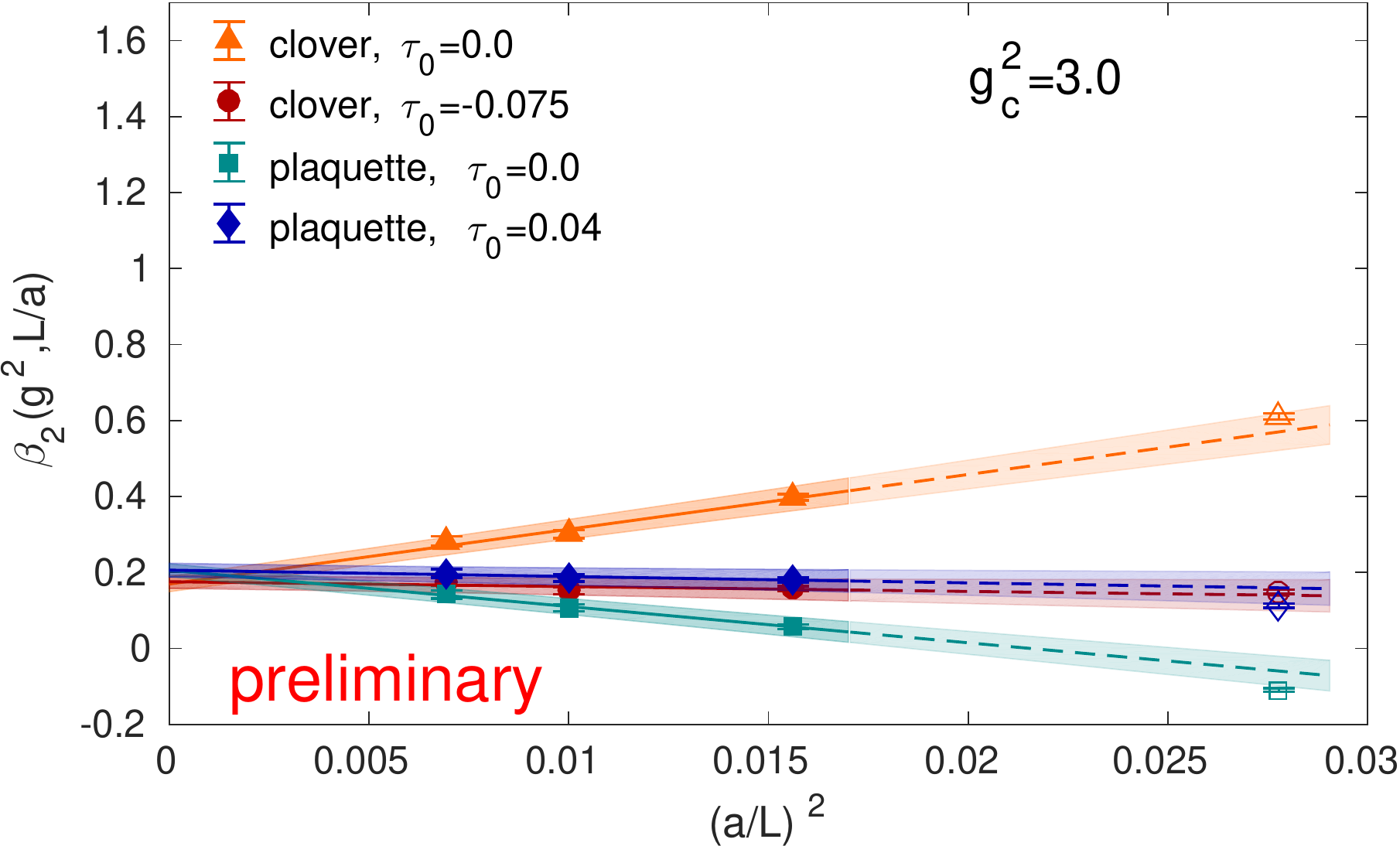}
  \includegraphics[height=0.2\textheight]{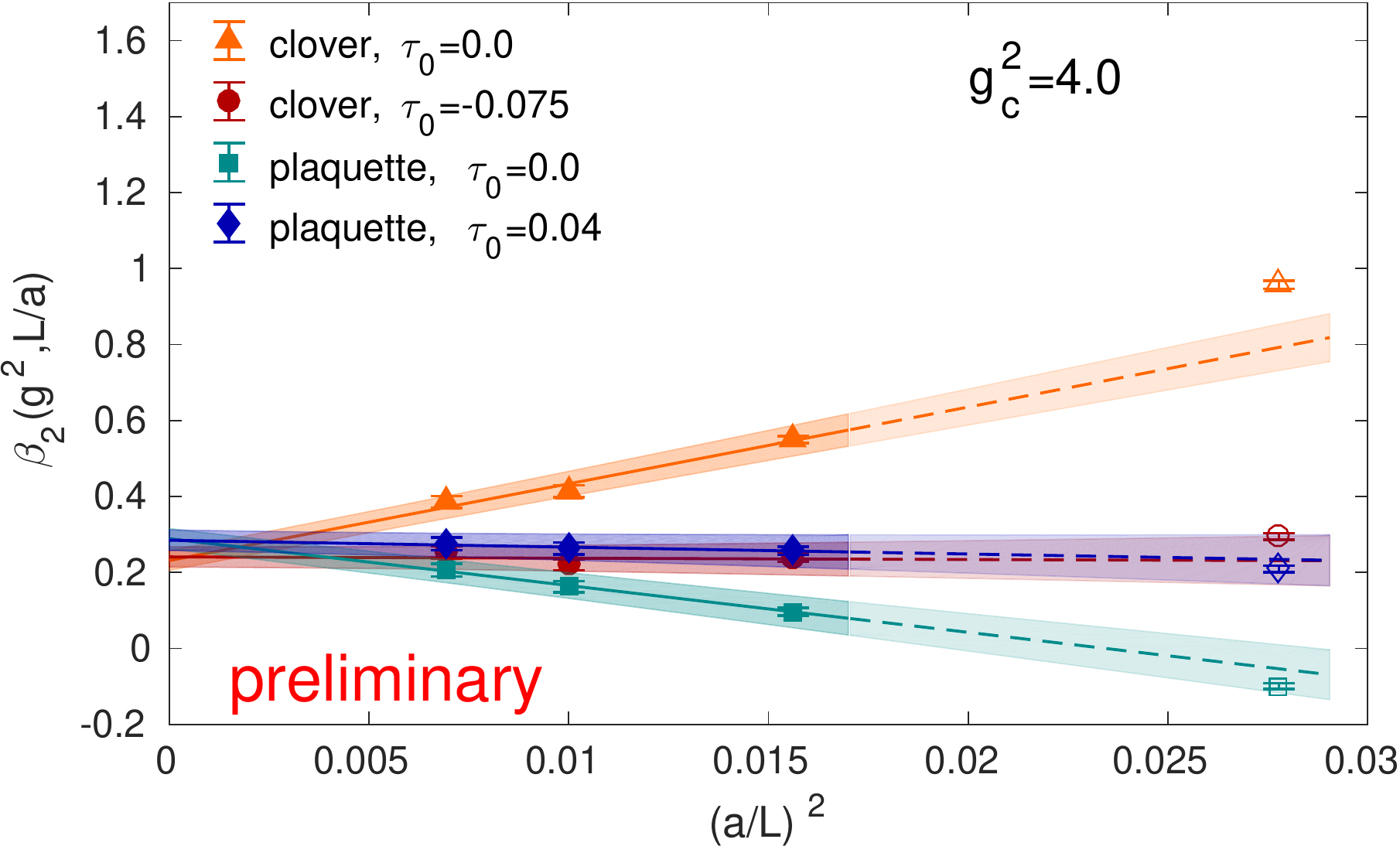}
   \includegraphics[height=0.2\textheight]{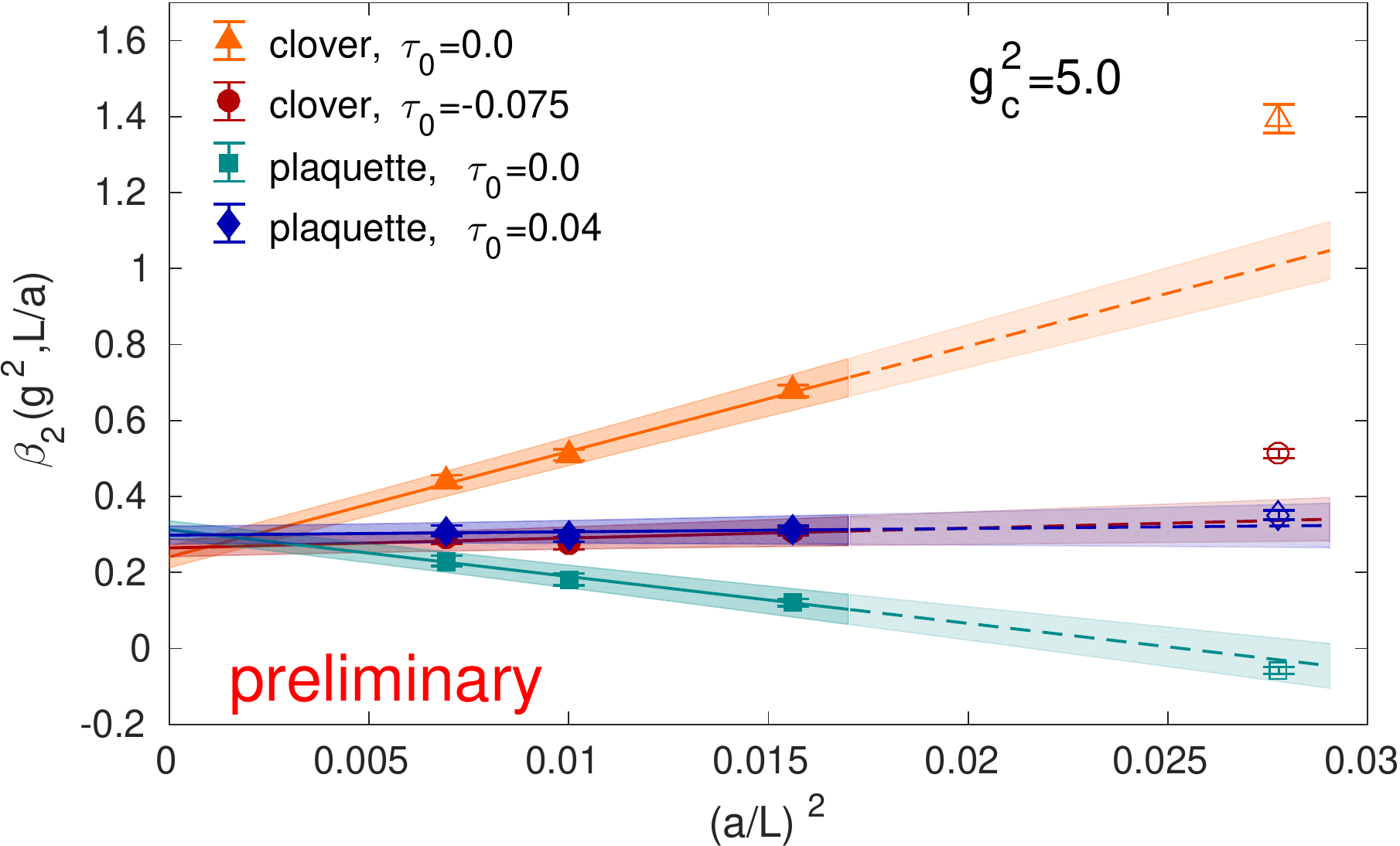}
  \includegraphics[height=0.2\textheight]{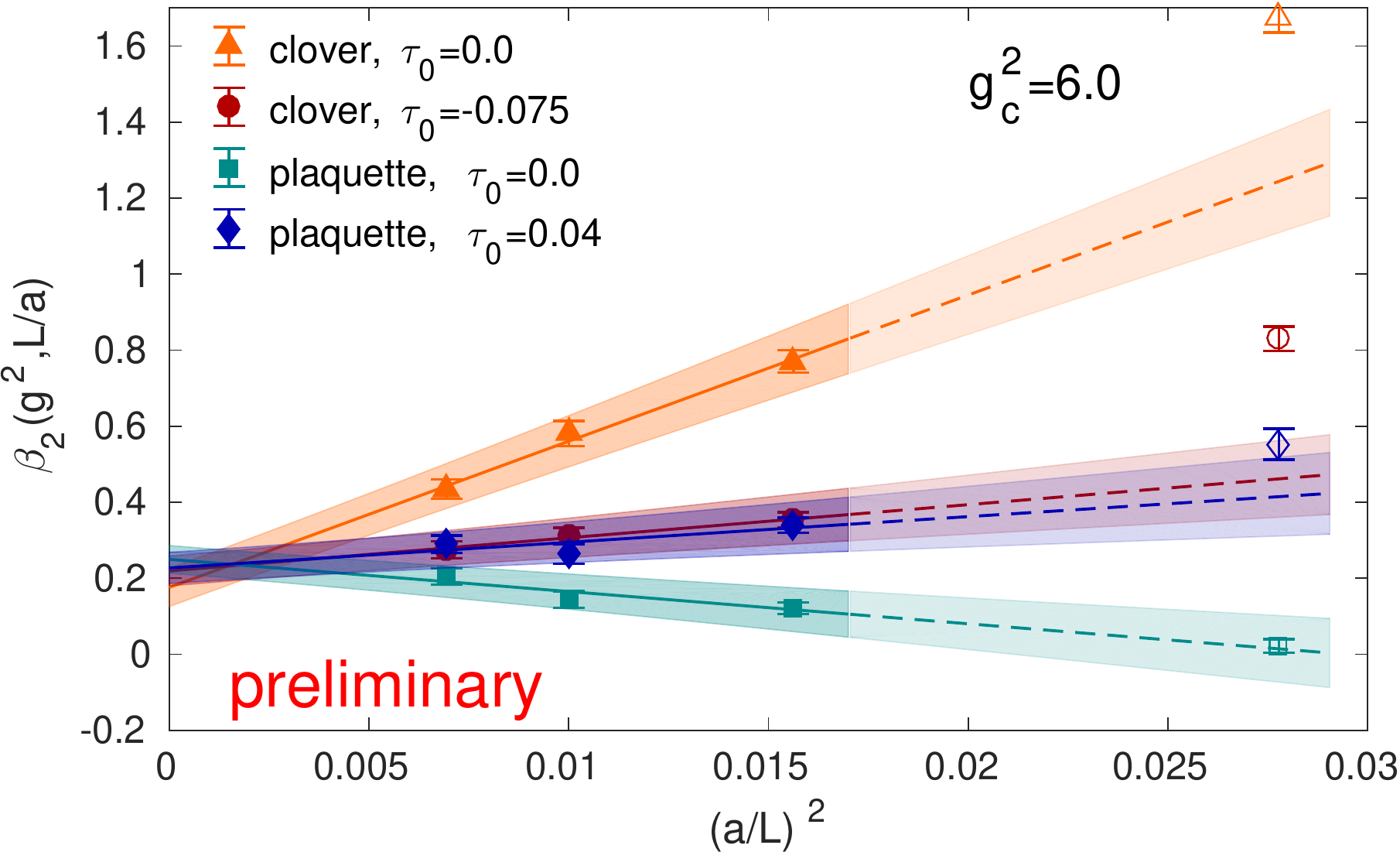}
  \caption{Continuum limit extrapolations in $N_f=10$ at $g^2_c=3.0$, 4.0, 5.0 and 6.0 in the $c=0.4$ gradient flow scheme. Each panel  shows the extrapolation in $(a/L)^2$ using the plaquette and clover discretizations both with and without  near-optimal $t$-shift values (0.04 for plaquette, -0.075 for clover).  The lines with errorband are linear fits to the data obtained from volumes $L/a \ge 8$ (filled symbols). We do not include the right-most data points (open symbols) corresponding to $L/a =6 \to 12$ matching but to guide eye extend our fit results to the right using dashed lines and a lighter shaded error band. Especially for the plaquette even those data points are close to the linear  fit lines, strengthening our confidence in the continuum extrapolation  based on volumes $L/a =8$ to $24$.}
  \label{fig:Nf10-continuum}
\end{figure}

\begin{figure}[tbh] 
  \centering
  \includegraphics[height=0.237\textheight]{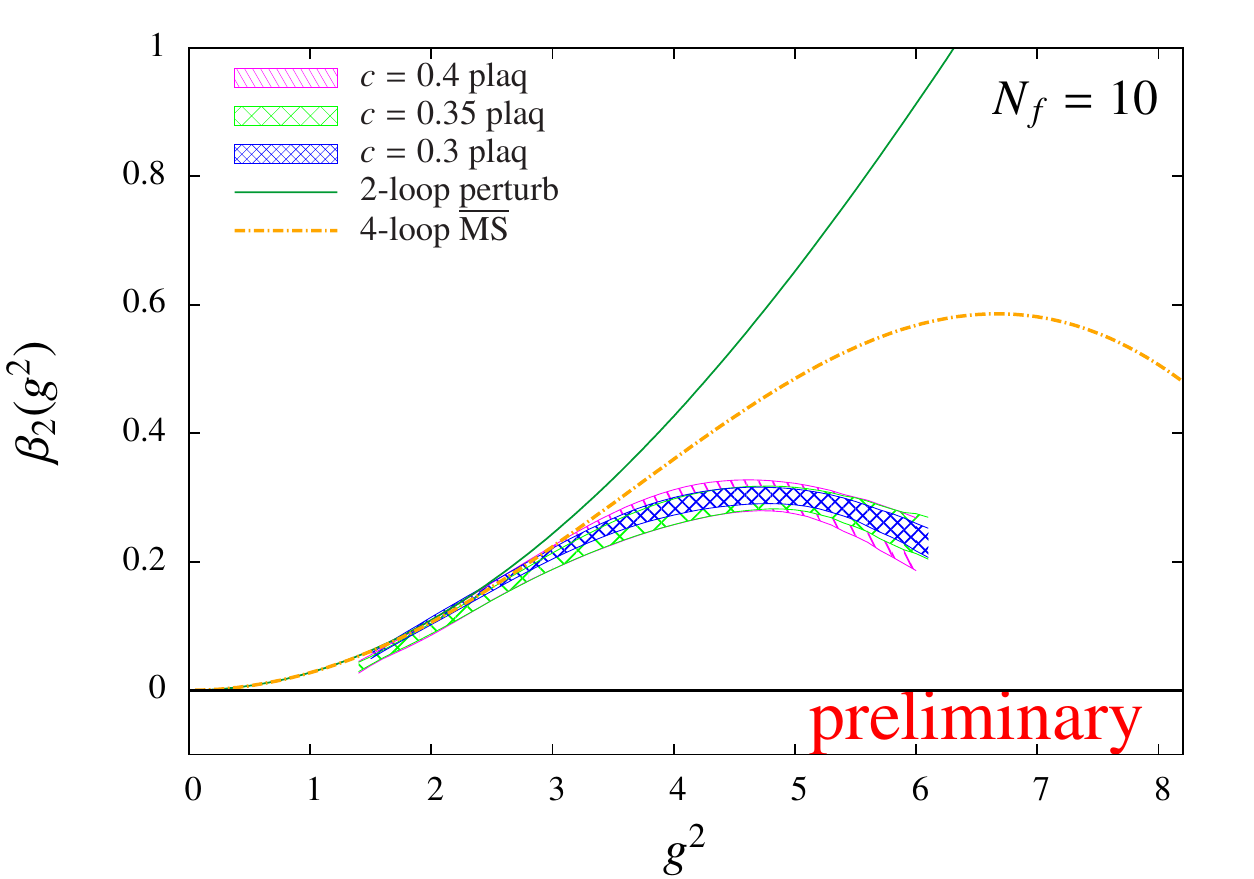}  
  \caption{Preliminary results for the $N_f=10$ step scaling function using MDW fermions. The continuum limit extrapolated predictions corresponding to $c=0.3$, 0.35 and 0.4 are consistent when using the plaquette discretization for the energy density and imply that the coupling is running slower than the 4-loop $\overline{MS}$ prediction.  We are not yet able  to verify the existence of the IRFP predicted in Ref.~\cite{Chiu:2016uui,Chiu:2017kza} but so far our  $\beta$-function agrees reasonably well with theirs.  } 
  \label{fig:Nf10-DW-plaq}
\end{figure}
As  mentioned in Sect.~\ref{sec:motivation}, the first domain wall simulation results of the 10-flavor gradient flow step scaling function  are reported in Refs.~\cite{Chiu:2016uui,Chiu:2017kza}. That work used  optimal DW fermions without smearing and considered volumes up to $L/a=32$. The result using $c=0.3$ shows that the step scaling function is below the 3-loop $\overline{MS}$ curve and indicates an  IRFP at $g^2_c\approx 7.0$. We do not have the computational resources to generate configurations at couplings $g^2_c \gtrsim 6.5$ as we would need $L_s\ge 32$ to keep the residual mass sufficiently small for $\beta \le 4.10$, and so far we did not reach an IRFP in our simulations. When comparing the plaquette and clover discretizations of the energy density, we observe a small but definite deviation between  the two discretizations for  $c=0.35$.  This disappears (within our statistical errors) when we increase  $c$ to 0.40,  as is shown in Fig.~\ref{fig:Nf10-DW-compare}.  The two discretizations are significantly different at $c=0.3$.

The  four panels of Fig.~\ref{fig:Nf10-continuum} show the $(a/L)^2$ extrapolations at $c=0.4$ for $g^2_c=3.0$, 4.0, 5.0, and 6.0. In all cases we consider both the plaquette and clover discretizations with  and without near-optimal $t$-shift. The $(a/L)^2$ extrapolations, depicted by solid  lines with error bands in Fig.~\ref{fig:Nf10-continuum} are based on volumes with $L/a \ge 8$.  Extending these fit lines to the right,  we  observe that for the plaquette discretization even the smallest $L/a=6 \to 12$ volumes are fairly close to  these linear fits. The four different extrapolations predict within errors consistent continuum limit values for $\beta_2(g^2_c)$. 
Since we use Wilson flow with Symanzik gauge action, the combination with plaquette discretization is expected to have smaller $\mathcal{O}(a^2)$ corrections~\cite{Ramos:2015baa}.  Fig.~\ref{fig:Nf10-DW-plaq} shows that the step scaling function obtained with the plaquette discretization is largely independent of $c$ when comparing $c=0.3$, 0.35, 0.4, a property that is not expected theoretically but seems to  hold also in other models. 
The continuum extrapolated step scaling function is  similar to the prediction of Refs.~\cite{Chiu:2016uui,Chiu:2017kza}. It follows the 2- and 4-loop perturbative curves up to $g^2_c\lesssim 3.0$ but turns away at larger couplings. Our results, shown in Fig.~\ref{fig:Nf10-DW-plaq}, explore the range for $g^2_c < 6.0$ but neither confirm nor discredit the  emergence of an IRFP.

\subsection{\texorpdfstring{Step scaling function with $N_f=12$ fundamental flavors}{Step scaling function with Nf=12 fundamental flavors}}\label{sec-Nf12}

\begin{figure}[thb] 
  \centering
  \begin{picture}(140,47)
    \put(0,0){\includegraphics[height=0.229\textheight]{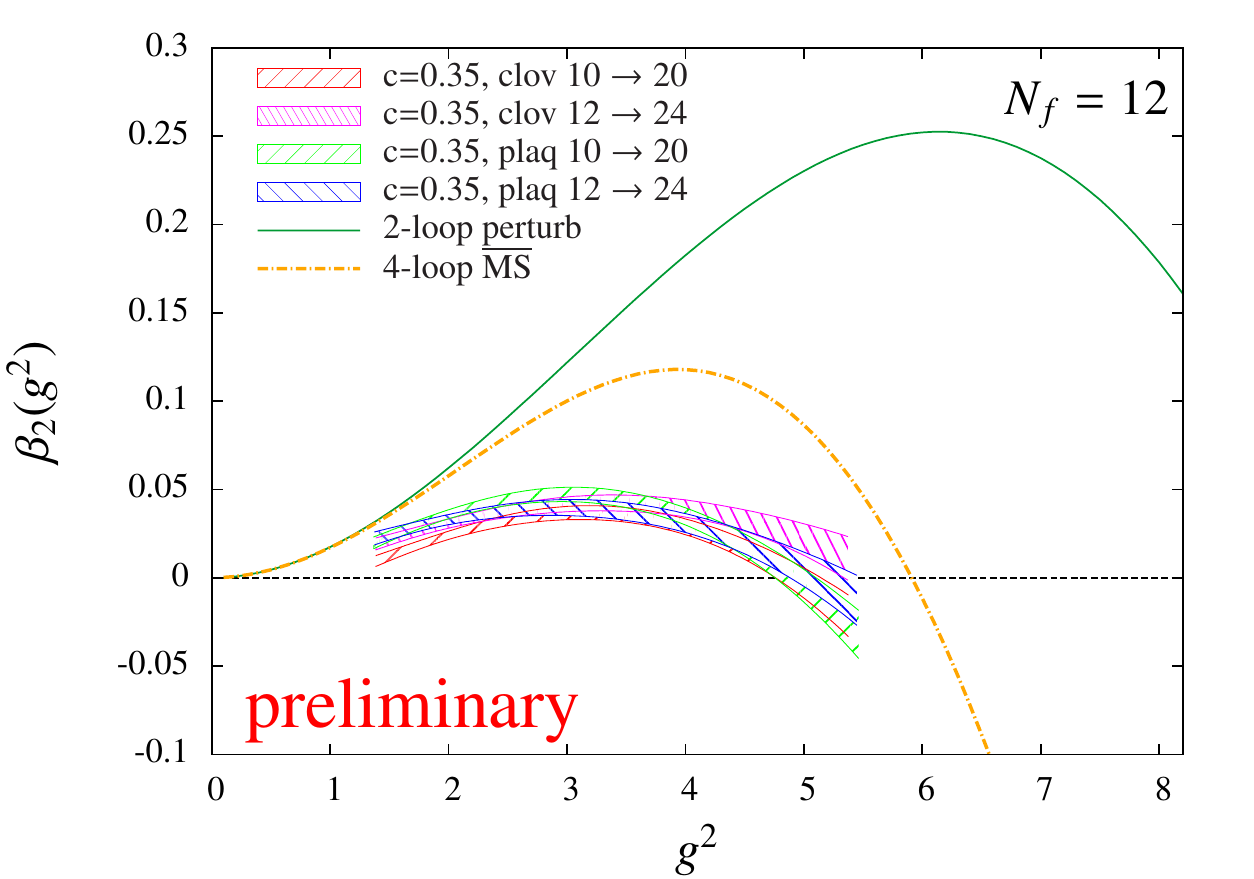}}    
    \put(69,1.3){\includegraphics[height=0.224\textheight]{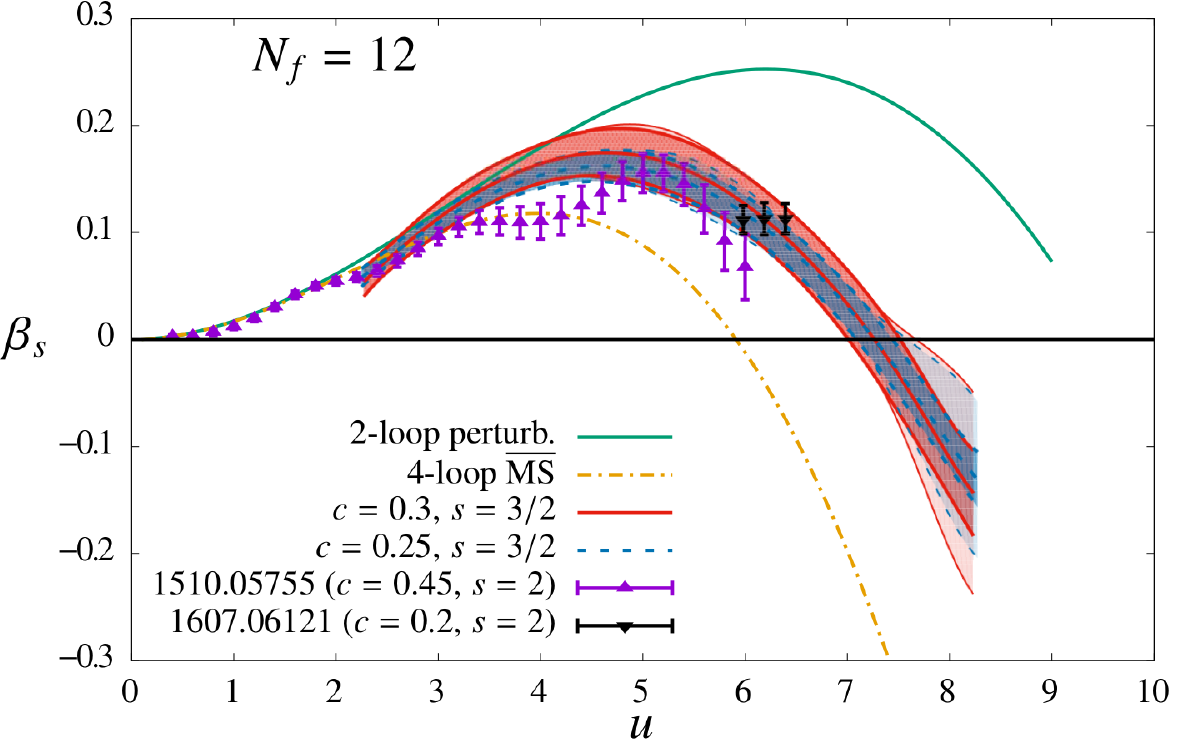}}    
  \put(115,42.8){\small arXiv:1610.10004}
  \end{picture}
  \caption{Left panel: Preliminary results for the $N_f=12$ step scaling function from our MDW fermion simulations. Using near-optimal $t$-shift (0.04  for the plaquette, -0.065  for the clover operator) the finite volume $\beta$ function of Eq.~(\ref{eq:beta_L}) is shown for $L/a=10\to 20$ and $12\to 24$. The dependence on $L$ is weak, hence the effect of the extrapolation to $L/a \to \infty$ is likely small. Larger $L/a=32$ volumes are in progress and will allow a standard infinite volume continuum extrapolation.  Right panel, reproduced from Ref.~\protect\cite{Hasenfratz:2016dou}: Comparison of the gradient flow step scaling functions  from Refs.~\protect\cite{Hasenfratz:2016dou} ($c = 0.25$  and $c = 0.3$, blue and red bands), \protect\cite{Lin:2015zpa} ($c = 0.45$, magenta symbols) and \protect\cite{Fodor:2016zil} ($c=0.2$, black symbols).  All three references use staggered fermions but given the different renormalization schemes and analysis details the results are in surprisingly good agreement. }  
  \label{fig:Nf12-DW}
\end{figure}
Given the tension between the step scaling function obtained for 12 flavors with staggered fermions and the outcome of
 the $N_f=10$ domain wall simulations presented above,  we have started to investigate the 12-flavor system with  DW  fermions to better  understand and hopefully resolve the conflict. 
The step scaling function with 12 flavors is small, making its non-perturbative determination  more challenging. While 5000 MDTU per configuration was sufficient to estimate $\beta_2(g^2_c)$ with 10 flavors, the same statistics is not sufficient with $N_f=12$. In addition we observe larger scaling violations  for the clover discretization when matching  $8 \to 16$ volumes  compared to our $N_f=10$ simulations.  Adding $L=32$ volumes will allow us to drop  the $L=8$ ensemble set or allow us to verify that all  four volume pairs are consistent with linear $(a/L)^2$  dependence. Comparing Wilson flow with Symanzik flow could also serve as a consistency check. At this point we  present only preliminary results. Figure~\ref{fig:Nf12-DW} shows the finite volume step scaling $\beta_2(L;g^2)$ as defined in Eq.~(\ref{eq:beta_L}) for the $10 \to 20$ and $12 \to 24$ volume pairs in the $c=0.35$ gradient flow scheme with both  plaquette and clover discretizations. The $t$-shift value was chosen to roughly minimize the volume dependence of $\beta_2(L;g^2)$, 0.04 for the plaquette discretization and -0.065 for the clover one. Once the $(a/L)^2$ extrapolation is taken, the value of the $t$-shift becomes irrelevant. In addition we use the same $t$-shift at every $g^2_c$,  proving further that its exact value is not important. Comparing the four $\beta_2(L;g^2)$ bands in Fig.~\ref{fig:Nf12-DW} shows that the continuum limit extrapolated step scaling function will  lie below the 4-loop $\overline{MS}$ prediction and also suggests an IRFP around $g^2_c=6.0$. Results with $c=0.4$ are similar. 

The step scaling function implied by our preliminary results are in  tension with the staggered fermion results that predict $\beta_s(g^2_c)$ increasing up to $g^2_c\approx 5$ before turning back and approaching  zero around $g^2_c=7.4$ as shown on the right panel of Fig.~\ref{fig:Nf12-DW}. However the DW step scaling function as presented in Fig.~\ref{fig:Nf12-DW} resolves the tension with the $N_f=10$ results shown in Fig.~\ref{fig:Nf10-DW-plaq}. Both the maximum and the implied zero of $\beta_s(g^2_c)$ occurs at stronger couplings with $N_f=12$ than $N_f=10$.

\section{Conclusion}\label{sec:conclusion}

Universality between continuum and staggered fermions at the Gaussian $g^2=0$ fixed point is subtle, but can be proven perturbatively.  This proof does not apply at a $g^2\ne 0$ non-trivial conformal IRFP  and a  simple 3-dimensional examples illustrates that universality can be broken near a conformal fixed point. Several existing lattice calculations point to tensions between staggered and other fermion formulations in 4-dimensional conformal or near-conformal systems. 

Motivated by these observations and the importance of conformal and  near-conformal models for  BSM phenomenology , we have initiated a study of the gradient flow step scaling function of $N_f=10$ and 12 fundamental flavors with the domain wall fermions. By comparing results with domain wall and staggered fermions we show that the two fermion formulations do not predict consistent results. 
Our results are preliminary. Increased statistics and simulations at additional coupling and volumes are needed to strengthen our conclusion.  If confirmed, results based on staggered fermion simulations of conformal systems or systems strongly influenced by a nearby conformal fixed point have to be taken with caution.
While these studies provide important information on conformal systems, their quantitative predictions might not be reliable.

\section*{Acknowledgments}
We are very grateful to Peter Boyle, Guido Cossu, Anontin Portelli, and Azusa Yamaguchi who develop the \texttt{Grid} software library providing the basis of this work and who assisted us in installing and running \texttt{Grid} on different architectures and computing centers. Computations for this work were carried out in part on facilities of the USQCD Collaboration, which are funded by the Office of Science of the U.S.~Department of Energy and the RMACC Summit supercomputer, which is supported by the National Science Foundation (awards ACI-1532235 and ACI-1532236), the University of Colorado Boulder, and Colorado State University. 
We thank  Fermilab,  Jefferson Lab, the University of Colorado Boulder, the NSF, and the U.S.~DOE for providing the facilities essential for the completion of this work.  A.H. acknowledges support by DOE grant 
DE-SC0010005 and C.R. by DOE grant DE-SC0015845.   This project has received funding from the European Union's Horizon 2020 research and innovation programme under the Marie Sk{\l}odowska-Curie grant agreement No 659322.

\bibliography{../General/BSM}

\end{document}